\documentclass[aps,prb,twocolumn,showpacs]{revtex4}

\usepackage{epsfig,graphics,graphicx}
\usepackage{color}
\usepackage{amssymb}
\usepackage{bookmath}

\definecolor{red}{rgb}{1,0,0}

\begin{document}


\title{Fermi-liquid effects in the Fulde-Ferrell-Larkin-Ovchinnikov state of 
two-dimensional $d$-wave superconductors}
\author{Anton B. Vorontsov}
\affiliation{Department of Physics and Astronomy, 
Louisiana State University, Baton Rouge, Louisiana 70803, USA}
\author{Matthias J. Graf}
\affiliation{Theoretical Division, 
Los Alamos National Laboratory, Los Alamos, New Mexico 87545, USA}

\date{\today  - LA-UR-06-2123}

\begin{abstract}
We study the effects of Fermi-liquid interactions on quasi-two-dimensional $d$-wave superconductors 
in a magnetic field. 
The phase diagram of the superconducting state, including the periodic 
Fulde-Ferrell-Larkin-Ovchinnikov (FFLO) state in high magnetic fields, is discussed for different 
strengths of quasiparticle many-body interactions within Landau's theory of Fermi liquids. 
Decreasing the Fermi-liquid parameter $F_0^a$ causes the magnetic spin susceptibility of
itinerant electrons to increase, which in turn leads to a reduction of the FFLO phase.
It is shown that a negative $F_0^a$ results in a first-order phase transition from the normal
to the uniform superconducting state in a finite temperature 
interval.
Finally, we discuss the thermodynamic implications of a first-order phase transition for CeCoIn$_5$.
\end{abstract}
\maketitle

\section{Introduction}

The coexistence of magnetism and superconductivity is generally thought
to be mutually exclusive. Recent discoveries of exotic superconductors,
which exhibit coexistence with ferromagnetism or antiferromagnetism, have challenged
this long standing belief.\cite{sax00,pfl01,par06} 
However, it has been known for some time that under certain circumstances
conventional superconducting order can coexist with paramagnetic order 
in high magnetic fields.
Since the 1960s this state of coexistence has become known as
Fulde-Ferrell-Larkin-Ovchinnikov (FFLO) state.\cite{ful64,lar64} 
More recently, it has been thought to be observed in the heavy-fermion
superconductor CeCoIn$_5$.\cite{bia03,rad03,kak05}

Here, we study the Pauli paramagnetic depairing effect of a magnetic field in 
a $d$-wave superconductor.
We use the weak-coupling Bardeen-Cooper-Schrieffer theory of 
superconductivity with quasiparticle interactions as described by Landau's 
theory of Fermi liquids.
The FFLO state of a spin-singlet superconductor in high magnetic fields
appears as a result of the intricate interplay of two effects: 
(1) loss of superconducting condensation energy 
in a magnetic field when Cooper pairs (with antiparallel spins) 
are depaired, and (2) gain of magnetic energy due to the Zeeman effect
by spin-polarizing quasiparticles.
As a result, in high fields a spatially nonuniform superconducting state 
coexists with pockets of spin-polarized quasiparticles localized 
at the zeros of the oscillating order parameter in real space.

We address the modifications of the FFLO state by Fermi-liquid (FL) interactions. 
Our analysis is for quasi-two-dimensional (quasi-2D) fermionic systems with a cylindrical 
Fermi surface and the magnetic field applied perpendicular to the axis of the cylinder.
In this geometry we can neglect any orbital effects on the superconducting condensate 
due to magnetic field. To further simplify our analysis, we restrict our study to
FFLO states with 1D spatial modulations of the order parameter
and neglect any low-temperature transition to a FFLO state with 2D modulations.\cite{shi98}

The effects of FL interactions in quasi-2D $s$-wave superconductors 
were first studied by Burkhardt and Rainer within quasiclassical theory.\cite{bur94} 
They reported a considerable change of the standard FFLO 
phase diagram when tuning the FL parameter $F^a_0$.
Since the underlying physics of Pauli depairing is the same for 
all spin-singlet superconductors, we expect similar new phenomena to occur for $d$-wave
pairing states.  However, we anticipate additional effects due to gap nodes and 
the associated spin-polarized nodal quasiparticles.

Here, we extend our earlier work\cite{vor05b} by including many-body interactions
in the form of FL effects, which enter the quasiclassical 
Eilenberger equation,\cite{eil68,lar68}
\be
[ i\vare_m \widehat{\tau}_3 - \widehat{v}_Z - \whs_\sm{FL} - \whDelta , \whg ] 
+ i\vv_f \cdot \grad \, \whg = 0 \ ,
\label{eq:eilFL}
\ee
with Matsubara frequency $\vare_m$ and quasiclassical Green's function $\whg$,
through the FL {\it dressed} Zeeman term, $\widehat{v}_Z$, and the FL self-energy,
$\whs_\sm{FL}$,\cite{ale85,ser83} 
\be
\widehat{v}_Z+\whs_\sm{FL} = \left(
\begin{array}{cc}
\vb \cdot \vsigma & 0 \\
0 & \vb \cdot \vsigma^* 
\end{array}
\right) \,, \quad \vb = \mu \vB_0/(1+F^a_0) + \vnu \,.
\ee
The Pauli matrices $\sigma_i$ describe the coupling of
quasiparticle spins to the
FL {\it dressed} external field $\vB_0/(1+F_0^a)$ and the internal exchange field $\vnu$.
The latter satisfies the self-consistency condition given by the spin part of the diagonal
component of $\whg$,
\be
\label{eq:nu}
\vnu(\vR) = A^a_0 \, T\sum_{\vare_m} \int d\hat{\vp}' \, 
 \vg(\vR, \hat{\vp}'; \vare_m) \,.
\ee
$A^a_0$ is the isotropic channel of the antisymmetric part of the Landau 
interaction $A^a(\hat{\vp}, \hat{\vp}')$.
The Landau parameter $A^a_0$ is related to the quasiparticle FL parameter 
$F^a_0$ through $A^a_0 = F^a_0/(1+F^a_0)$.
$\mu = (g/2)|\mu_B|$ is the magnetic moment of an electron.
The $g$ factor of a free electron is $g=2$.
Here $g$ is a free parameter.

\section{Results and Discussion}

\begin{figure}[t]
\centerline{\includegraphics[height=55mm]{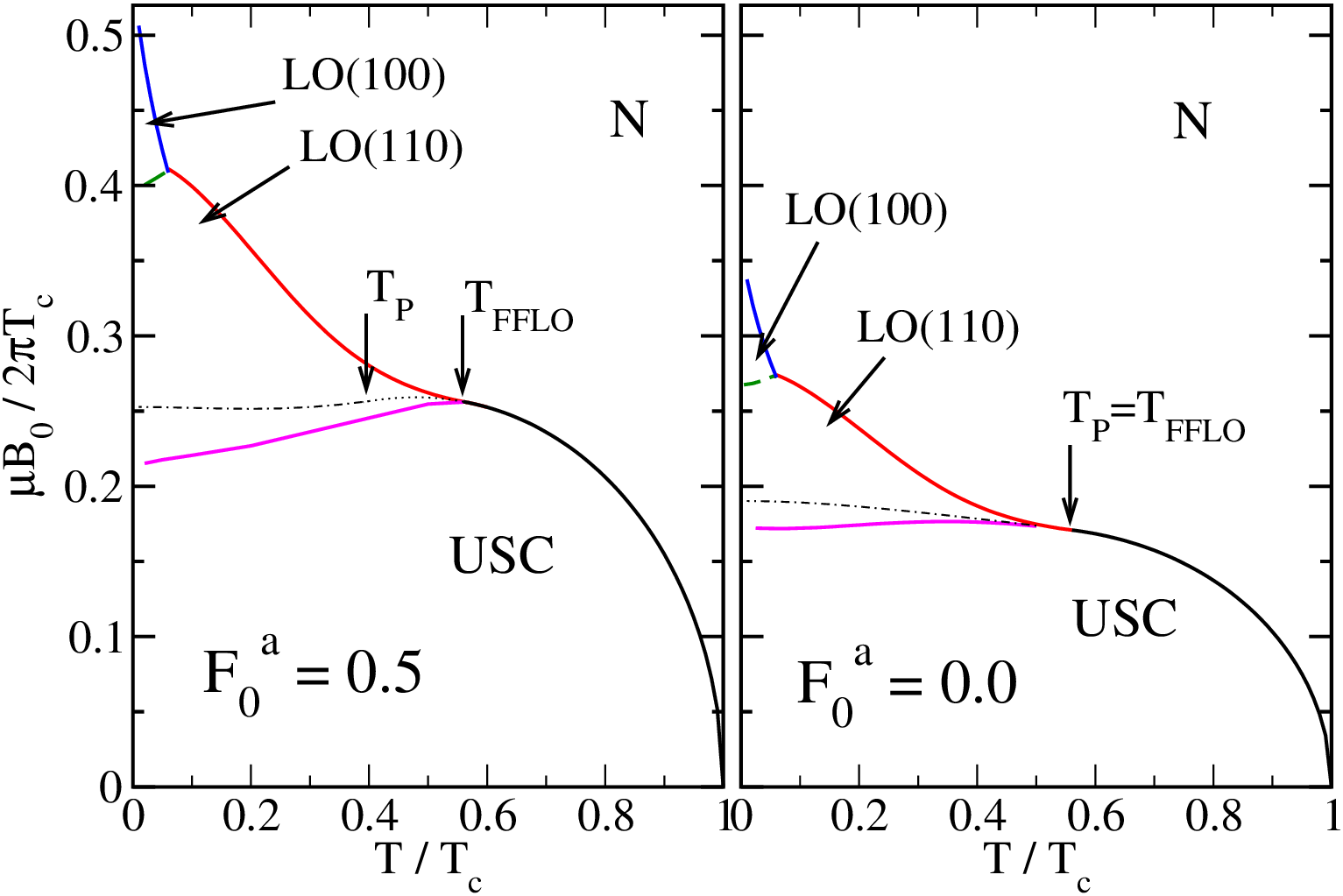}}
\caption{\label{fig:FL_0.5_0.0} 
(Color online) The phase diagram of a 2D $d_{x^2-y^2}$-wave superconductor for 
$F_0^a = 0.5$ (left panel) and $F_0^a = 0.0$ (right panel). 
For positive $F_0^a$ the LO state is stabilized over a wider range of fields.
Note that the energetically unphysical Pauli-limited transition is 
first order (dot-dashed black line) at low temperatures, 
$0<T<T_P<T_{FFLO} \approx 0.56 T_c$. 
Above $T_P \approx 0.4 T_c$ the instability would become second order (dotted line).
Without Fermi liquid effects $T_P=T_{FFLO}$.
At the lower critical field $B_{c1}$ a
LO state with modulation $\vq$ along nodal directions  $(110)$
is stablized (solid magenta line). 
At even higher fields a LO state with $\vq \parallel (100)$ becomes stable (dashed green line).
}
\end{figure}

\begin{figure}[t]
\centerline{\includegraphics[height=55mm]{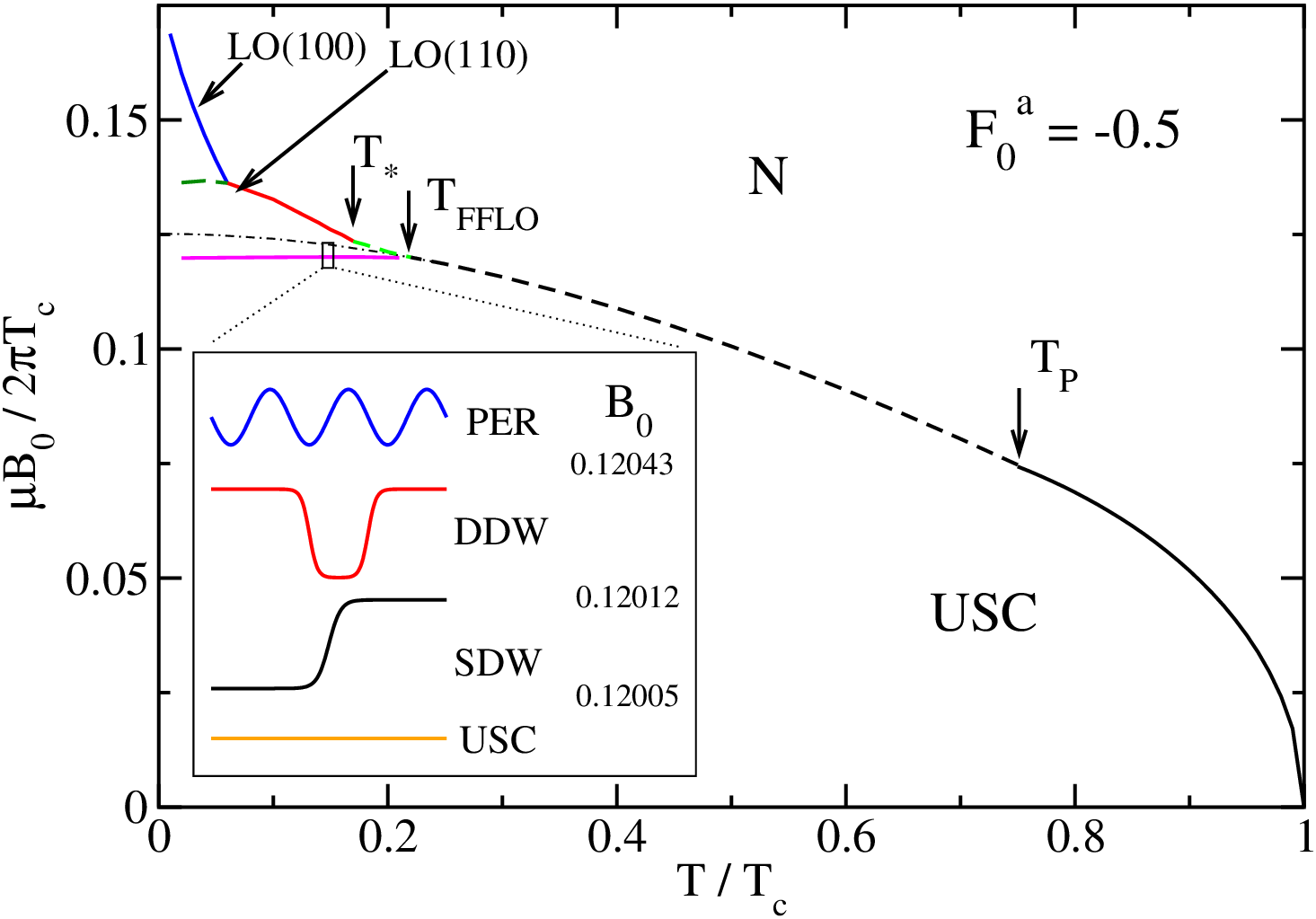}}
\caption{\label{fig:FL_F-0.5} 
(Color online)
The phase diagram of a 2D $d$-wave superconductor for $F_0^a = -0.5$
with $T_{FFLO} \ll T_P \approx 0.75 T_c$.
Note the first-order LO-N transition between $T_* < T  < T_{FFLO}$.
Inset: Sketch of evolution of the order parameter near $B_{c1}$ transition from uniform (USC) 
to periodic (PER) LO solution at $T/T_c=0.15$. 
In a narrow wedge of magnetic fields single (SDW) and double (DDW) domain wall 
solutions are favored over either USC or PER states. 
}
\end{figure}

In Figs.~\ref{fig:FL_0.5_0.0} and \ref{fig:FL_F-0.5} we show the computed phase diagrams 
of a 2D $d$-wave superconductor for three different strengths of the FL parameter 
$F^a_0$ ranging from negative to positive.
The evolution of the $d$-wave phase diagram with $F_0^a$ is similar to the 
$s$-wave case.\cite{bur94} 

We determine the order of a phase transition by calculating 
the jump in the spin magnetization (density
of the magnetic moment),\cite{ale85}
\be
\vM(\vR) = \frac{2\mu N_\sm{F}}{1+F_0^a} 
\left( \mu \vB_0 - 
 T\sum_{\vare_m} \int d\hat{\vp}' \, \vg(\vR, \hat{\vp}'; \vare_m) 
\right) \,,
\label{eq:mag} 
\ee
across the transition line. Simultaneously, we check
it by directly evaluating the free energy.
A discontinuity of the magnetization, $\vM ={\partial (F/V)}/{\partial \vB}$, 
defines a
first-order phase transition, while 
a kink in $\vM$ (discontinuity in the susceptibility)
defines a second-order transition.  

We adopt the following notation for drawing transition lines: 
second-order transitions have solid lines, while
first-order transitions have dashed lines. 
For comparison, the unphysical part of the normal (N) 
to uniform superconducting (USC) state transition (Pauli limited)
is shown by a thin dot-dashed line inside the Larkin-Ovchinnikov (LO) phase,
as well as the corresponding second-order phase transition between USC and LO 
phases (solid magenta line).

\paragraph{Second-order transition line at $B_{c2}$:}
First, we consider the second-order instability line of the upper critical field
$B_{c2}$ from the N state into 
the USC state or from the N state into the spatially 
nonuniform (periodic) FFLO state with an order parameter $\Delta(\vR) \sim \exp(i\vq\cdot\vR)$ 
or $\Delta(\vR) \sim \cos \,\vq\cdot\vR$.
The phase transition can be obtained by linearizing the Eilenberger equation in $\Delta$.
Near a second order transition $\vnu$ is zero in linear order. Thus, 
the linearized gap equation for $\Delta$ is identical 
to that of the $F_0^a=0$ case if one replaces $B_0$ with $B_0/(1+F_0^a)$.
So one obtains the second order normal-state instability line 
from the known solution\cite{shi98,shi97,mak96,yan98,vor05b} by simple scaling,
$ \mu B_{c2}(T;F_0^a) = \mu B_{c2}(T;0) (1+F_0^a)$.

\paragraph{First-order transition line at $B_{c2}$:}
For first-order transitions, we must solve the general expressions (\ref{eq:eilFL}) - 
(\ref{eq:nu}), which are nonlinear in the mean fields, and calculate
the corresponding Green's functions and free energy. 
However, the calculation of the Pauli-limited 
transition line from the normal into the uniform state 
is straightforward, since we know already the general form of $\whg$
for a uniform superconductor.\cite{vor05b}
At $T=0$, the free energy density can be expressed in a very
intuitive way, 
$\Del F/V =
-{1\over 2} N_\sm{F} \langle |\Delta(\hat{\vp})|^2 \rangle_\sm{FS}
+ {1\over 2} \Del \vM \cdot \vB_0 $, 
similar to the result by Clogston,\cite{clo62} 
where $\Del \vM = \vM_N - \vM$ and $N_\sm{F}$ is the density of states per spin
at the Fermi energy. It follows
that a gain in condensation energy happens at the expense of 
the magnetic energy of the spin-polarized quasiparticles, which is proportional to 
the difference of the spin magnetization between the N and USC state.
In the $s$-wave superconductor in the absence of the Meissner effect, 
the electron magnetization $\vM$ vanishes at $T=0$, and the Pauli limited field is
$\mu B_P = \sqrt{ {1\over 2}(1+F_0^a) \langle |\Delta(\hat{\vp};B_P)|^2 \rangle_\sm{FS}}$.
For $d$-wave, $\vM(T=0)$ is nonzero due to nodal quasiparticles,
but is reduced from the normal-state magnetization by a fraction $p=|\Del \vM|/|\vM_N|$. 
Then the right-hand side of the previous equation needs to be divided by $\sqrt{p}$.

If many-body interactions, like FL effects, are considered in the N state, 
then the spin magnetization of an isotropic Pauli paramagnet is given by
$\vM_N = \chi_N \, \vB_0 = 2\mu^2 N_\sm{F} \vB_0 / (1+F_0^a)$.
Thus, for positive FL parameters $F_0^a$ the normal-state susceptibility $\chi_N$ 
is suppressed compared to a noninteracting Fermi gas, while for negative $F_0^a$ it
is enhanced. The FFLO state, as well as the USC state, become stable 
in a wider range of magnetic fields for positive $F_0^a$ (Fig.~\ref{fig:FL_0.5_0.0}), 
and in a smaller region for negative $F_0^a$ (Fig.~\ref{fig:FL_F-0.5}).  
This happens because pockets of polarized electrons do not gain as much in
magnetic energy for $F_0^a>0$ as for $F_0^a=0$, and the opposite happens for $F_0^a<0$. 
The FFLO state exists for any $F^a_0>0$, but disappears at some critical negative value,
when the upper critical field $B_{c2}$ of the FFLO state, 
$B_{c2}(T;F_0^a) = B_{c2}(T;0) (1+F_0^a)$, drops below the Pauli-limited field,
$B_{P}(T;F_0^a) \approx B_{P}(T;0) \sqrt{1+F_0^a}$.
At that point the FFLO state 
becomes unstable against the USC state for any field and temperature. 
The numerically determined critical value is
$F_0^a \approx -0.765$ for a $d$-wave superconductor, which is 
lower than that for an $s$-wave superconductor, 
$F_0^a=-0.5$.\cite{bur94} 

We see that the FFLO state in a 2D $d$-wave superconductor is more stable
against FL effects compared to the $s$-wave case discussed in detail by
Burkhardt and Rainer.\cite{bur94} 
Again, this is not completely unexpected, because spin-polarized nodal quasiparticles
can gain magnetic energy without the breaking of
Cooper pairs, which is unavoidable in a fully gapped superconductor. 

Further, negative values of $F_0^a$ make part of 
the LO-N transition, $T_* < T < T_{FFLO}$, to be 
first order (Fig.~\ref{fig:FL_F-0.5}). 
This is in qualitative agreement with the $s$-wave case.\cite{bur94} 
However, in contrast to the $s$-wave case we failed to detect any 
first-order transition inside the LO phase 
between phases with different periods $\vq$.

\paragraph{Second-order transition line at $B_{c1}$:}
Fig.~\ref{fig:FL_F-0.5} also shows details of the lower critical transition
$B_{c1}$ from the USC state to the periodic LO state. We calculate the free energy 
as a function of the field for four different types of order parameters. 
We find successive transitions from the USC solution to the single-domain wall (SDW) and 
then to the double-domain wall (DDW) and next to the periodic solution (PER) in a 
thin but finite wedge of magnetic fields. For $T/T_c=0.15$, 
we sketch in the inset of Fig.~\ref{fig:FL_F-0.5} 
the different energetically favorable solutions with their respective transitions in
magnetic field.
Although the transition from the USC to SDW state is continuous, there is a sequence 
of transitions, as domain walls enter the bulk one by one with increasing 
field. 

\begin{figure}[th]
\centerline{\includegraphics[height=60mm]{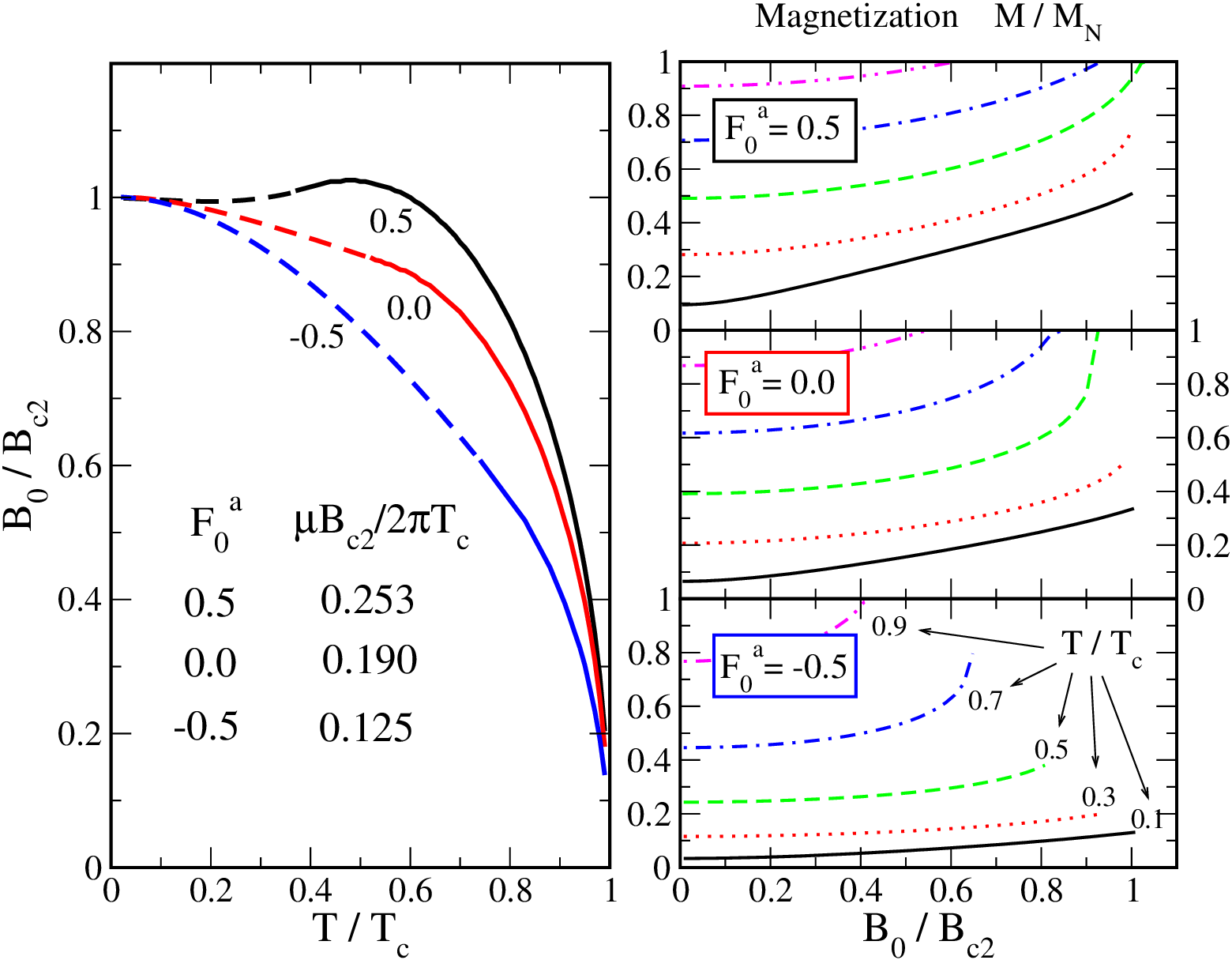}}
\caption{\label{fig:hc2} 
(Color online)
Left panel: The Pauli-limited upper critical field for Fermi-liquid parameters 
$F_0^a=\{0.5, 0.0, -0.5\}$.
Right panel: Normalized uniform magnetization vs.\ field for temperatures 
$T/T_c=\{0.1, 0.3, 0.5, 0.7, 0.9\}$ as labeled in the bottom window.
}
\end{figure}

\paragraph{The USC-N transition:} 
Finally, in Fig.~\ref{fig:hc2} we show the normalized 
Pauli limited transition between the normal and 
uniform superconducting state for several Fermi-liquid parameters. 
The break in line from solid to dashed indicates, as before, the change from
a second to first-order transition.
On the right side of Fig.~\ref{fig:hc2} we show the normalized magnetization 
in the USC state as a function of field. Note that the magnetization jump, 
when crossing into the normal state, is larger for negative Fermi-liquid 
parameters.

\section{Thermodynamic implications}

So far we discussed specific results for the phase diagram of a quasi-2D Fermi liquid
model in the superconducting state. As we have seen, the existence and extent of a first-order
phase transition between the normal and superconducting state - USC or FFLO - at lower temperatures
can be modeled by invoking a negative Fermi liquid parameter $F_0^a$.
On the other hand, in three dimensional superconductors of type-II 
the first-order transition line can be modified by a combination of Zeeman and orbital depairing.
Generally, the existence of a first-order phase transition between the normal and superconducting state
is an anomalous phenomenon for strong type-II superconductors. 
Since the first-order transition {\it is} seen in CeCoIn$_5$, we want to address
the thermodynamic constraints imposed along such transition independent of a specific
model. This is an important question that needs to be addressed if one wants to compare
theory with experiments.

Although most experiments report phase diagrams for CeCoIn$_5$ in fair
agreement with each other (see Refs.\ \onlinecite{tay02,bia03,kak05,mit06,Curro06}), 
there is a noticeable variation in the position and sharpness of the 
first-order transition between the normal and superconducting state.
For that reason, we check the internal consistency of independent experiments
by a thermodynamic analysis.
Along the first-order transition line the free energy is continuous 
and results in a generalized Clausius-Clapeyron equation, 
$\frac{d B_{P}}{d T} = -\frac{ \Del S }{ V \Del M }$,
which relates the jumps in entropy, $\Del S = S_N - S$, and magnetization,
$\Del M = M_N - M$, and volume $V$.
If the magnetization in the superconducting state is reduced by a fraction $0<p<1$ from 
$M_N$, then $\Del S/V = - p M_N d B_{P}/ d T$.
Consequently, the latent heat associated with this transition is
$Q=T \Del S = p T M_N |d B_{P}/ d T|$.

The experiments by Bianchi et al.\cite{bia03} show a
value of $\Del S/V_{\rm mol} \approx 200$
mJ/(mol\ K) at $T \approx 0.5$K,
and a measured slope of the upper critical field of
$d B_{P}/dT \approx -1.5$ T/K at $T\approx 0.5$ K.\cite{tay02,bia03}
Whereas Tayama et al.\cite{tay02} reports a
magnetization jump of
$\Del M \approx 0.1\, M_N \approx 80$ mJ/(mol\ T) at $T \approx 0.45$ K.
It is obvious that the agreement between the ratio of the discontinuities,
$\Del S/(V  \Del M) \sim 2.5$ T/K, 
and the slope of $B_{c2}$ is poor at $T\approx 0.5$ K 
(a deviation of $\sim 70$\%), despite good overall
agreement between both phase transition lines.
The origin of this inconsistency is poorly understood, but might
be related to the nature of the localized $f$ electrons and their contributions
to the magnetization.  
Further entropy and magnetization studies are needed to resolve this open problem.

\section{Conclusions}

We studied the superconducting phase diagram of quasi-2D $d$-wave superconductors
in the presence of Fermi-liquid effects in high magnetic fields.  We found that 
a negative Fermi-liquid parameter $F_0^a$ increases the gain in magnetic energy,
while at the same time it reduces the available phase space of a FFLO state. The 
uniform superconducting and periodic FFLO state are competing and at a critical 
Fermi-liquid parameter $F_0^a\approx -0.765$ the FFLO state is completely suppressed.

We note that in order to explain the high-field phase diagram of the CeCoIn$_5$ superconductor 
in terms of an FFLO state one needs to go beyond the simplistic Fermi-liquid picture 
considered here. 
While we find that the inclusion of Fermi-liquid effects considerably changes 
the phase diagram, the changes are not consistent with the experimental findings. 
For example, (1) the magnitude of the calculated critical temperature $T_{P}$, 
where the Pauli-limited upper critical transition changes from second to first order
between the uniform superconducting  and normal state, and (2) the corresponding magnetization
jump are much larger than those seen in experiment. 
We also note  that the shape of the transition lines of the calculated FFLO state is 
qualitatively different from experiments.
It is not unreasonable to expect that some of those discrepancies may be overcome by including 
the effects of impurity scattering,\cite{ada03} antiferromagnetic spin fluctuations, local
magnetic moments, or orbital effects.

Finally, our analysis of the first-order transition puts stringent constraints on thermodynamic
properties and reveals a significant discrepancy between specific heat and magnetization
measurements that requires further studies.

\section{Acknowledgments}

We thank I. Vekhter, C. Capan, R. Movshovich, J. Thompson, and L. Bulaevskii for helpful discussions. 
We are grateful for the parallel computing resources at T-CNLS and the 
IC supercomputer facilities at LANL.
A. B. V. received funding from the Louisiana Board of Regents for this 
research.  M. J. G. was supported by the U.S. DOE at Los Alamos National Laboratory 
under the auspices of the NNSA under Contract No. DE-AC52-06NA25396.

\bibliographystyle{apsrev}

\end{document}